\documentclass[11pt,reqno]{amsart}
\usepackage{fullpage}
\usepackage{mathrsfs,amssymb,graphicx,verbatim,amsmath,amsfonts}
\usepackage{paralist}
\usepackage{esint}
\usepackage{fourier}
\usepackage[breaklinks,pdfstartview=FitH]{hyperref}
\usepackage{upgreek}
\usepackage[mathscr]{euscript}
\usepackage[T1]{fontenc}

\makeatletter
\def\resetMathstrut@{%
  \setbox\z@\hbox{%
    \mathchardef\@tempa\mathcode`\(\relax
    \def\@tempb##1"##2##3{\the\textfont"##3\char"}%
    \expandafter\@tempb\meaning\@tempa \relax
  }%
  \ht\Mathstrutbox@1.2\ht\z@ \dp\Mathstrutbox@1.2\dp\z@
}
\makeatother

\usepackage{graphicx}
\usepackage{sidecap}
\usepackage[all]{xy}
\usepackage[font=small,labelfont=bf,labelsep=endash]{caption}
\usepackage{enumitem}

\renewcommand{\eta}{\upeta}

\renewcommand{\epsilon}{\upepsilon}

\newcommand{\FF}{\mathcal{F}}

\renewcommand{\chi}{\upchi}

\renewcommand{\le}{\leqslant}
\renewcommand{\ge}{\geqslant}

\renewcommand{\gamma}{\upgamma}
\renewcommand{\lambda}{\uplambda}
\renewcommand{\alpha}{\upalpha}
\renewcommand{\delta}{\updelta}
\renewcommand{\beta}{\upbeta}
\renewcommand{\omega}{\upomega}
\renewcommand{\nu}{\upnu}
\renewcommand{\mu}{\upmu}
\renewcommand{\psi}{\uppsi}
\renewcommand{\phi}{\upphi}
\renewcommand{\rho}{\uprho}
\renewcommand{\SS}{\mathbb{S}}
\renewcommand{\kappa}{\upkappa}

\renewcommand{\tau}{\uptau}

\newcommand{\ud}[0]{\,\mathrm{d}}

\makeatletter
\def\moverlay{\mathpalette\mov@rlay}
\def\mov@rlay#1#2{\leavevmode\vtop{%
   \baselineskip\z@skip \lineskiplimit-\maxdimen
   \ialign{\hfil$\m@th#1##$\hfil\cr#2\crcr}}}
\newcommand{\charfusion}[3][\mathord]{
    #1{\ifx#1\mathop\vphantom{#2}\fi
        \mathpalette\mov@rlay{#2\cr#3}
      }
    \ifx#1\mathop\expandafter\displaylimits\fi}
\makeatother

\newcommand{\sign}{\mathrm{sign}}

\newcommand{\n}{\{1,\ldots,n\}}

\renewcommand{\t}{\uptau}

\newcommand{\PP}{{\mathbb P}}

\newcommand{\eps}{\varepsilon}
\newcommand{\R}{\mathbb R}
\newcommand{\RR}{\mathbb R}

\newtheorem{theorem}{Theorem}

\theoremstyle{remark}
\newtheorem{remark}[theorem]{Remark}

\renewcommand{\pi}{\uppi}
\renewcommand{\zeta}{\upzeta}

\renewcommand{\sigma}{\upsigma}

\newcommand{\N}{\mathbb N}

\newcommand{\eqdef}{\stackrel{\mathrm{def}}{=}}

\newcommand{\A}{\mathsf{A}}
\newcommand{\EE}{\mathsf{E}}

\begin{document}

\title{Krivine diffusions attain the Goemans--Williamson approximation ratio}


\author{Ronen Eldan}

\author{Assaf Naor}

\thanks{R.~E. is supported by a European Reseach Council Starting Grant (ERC StG) and by the Israel Science Foundation. }
\thanks{A.~N. was supported by the Packard Foundation and the Simons Foundation.}
\thanks{The research presented
here was conducted under the auspices of the Simons Algorithms and Geometry (A\&G) Think Tank.}

\maketitle


\begin{abstract} Answering a question of Abbasi-Zadeh, Bansal, Guruganesh, Nikolov, Schwartz and Singh (2018), we prove the existence of a slowed-down sticky Brownian motion whose induced rounding  for MAXCUT  attains the Goemans--Williamson approximation ratio. This is an especially simple particular case of the general rounding framework of Krivine diffusions that we  investigate elsewhere.
\end{abstract}


\subsubsection*{Bibliographical comment} This text amounts to an extraction of a subsection from our preprint~\cite{EN14}, with a modicum of background details added so as to make it self-contained. \cite{EN14} is an outgrowth of a project that we have been pusuing over the past 5 years to understand a framework, which we call {\em Krivine diffusions}, to consistently ``round'' high-dimensional vectors to signs, i.e., to elements of $\{-1,1\}$. We decided to post the present note separately, prior to the publication of the more extensive manuscript~\cite{EN14} (at which time~\cite{EN14} will subsume the present note), because recently a special case of the approach of~\cite{EN14} arose in~\cite{ABGNSS18}, and we realized that even this (much simpler) setting of~\cite{EN14} resolves one of the main open questions of~\cite{ABGNSS18}. The present extract from~\cite{EN14} establishes this fact in a self-contained manner that can be read as a standalone article without waiting for the appearance of~\cite{EN14}, though~\cite{EN14}  situates the ensuing reasoning within proper context and derives more delicate statements in other settings of interest (notably for the Grothendieck constant, which entails treating cancellations that do not arise in the present setting).

\section{Introduction}

Throughout what follows, we fix $d\in \N$ and let $\langle\cdot,\cdot\rangle:\R^d\to \R$ be the standard scalar product on $\R^d$, with the associated unit sphere  denoted by $\mathbb{S}^{d-1}=\{x\in \R^d:\ \langle x,x\rangle=1\}$.

Suppose that $n\in \N$ and $x_1,\ldots,x_n\in \SS^{d-1}$. It is a ubiquitous problem in several areas to meaningfully associate to these vectors signs $\sigma_1,\ldots,\sigma_n\in \{-1,1\}$. The interpretation of the term ``meaningfully'' here is problem-specific, and it encompasses issues that may include the efficiency of determining these signs.  This is of central importance to approximation algorithms, when $x_1,\ldots,x_n$ are an output of a continuous relaxation of a combinatorial optimization problem. Grothendieck famously broached~\cite{Gro53} this matter for (formidable) analytic applications (see also the work of Lindenstrauss and Pe{\l}czynski~\cite{LP68} and the survey~\cite{Pis12}), and it occurs  in quantum information theory; see e.g. the work of Tsirelson~\cite{Tsi85}.

The above generic ``rounding'' problem has the following seemingly na\"ive solution: Simply output the ``first bit'' (namely, the sign of the first entry) of each of these vectors, perhaps after first applying a suitable transformation, such as a uniformly random rotation as in the work of Goemans and Williamson~\cite{GW95} or a more sophisticated ``preprocessing'' as in the work of Krivine~\cite{Kri77}. Despite its simplistic appearance, this procedure is of major importance to algorithm design, as demonstrated decisively in the seminal work~\cite{GW95} and in numerous subsequent investigations.

Superior performance of the above rounding task can sometimes be achieved by associating a sign to a  vector in a manner that is more sophisticated than considering its first bit after preprocessing. This may be somewhat  nonintuitive at times, e.g.~Krivine~\cite{Kri77} stated that this is not so for the Grothendieck problem (to be described below), but~\cite{BMMN-grothendieck} established that Krivine's conjecture has a negative answer (note that despite the fact that for historical reasons this is formally described as a ``negative answer'' to a conjecture, it entails the design of a better rounding method, so it is in fact a positive result). Such higher-dimensional ``Krivine rounding schemes'' are shown to be optimal for the Grothendieck problem in~\cite{NR14}, and this is the reason for the nomenclature that we choose below to describe a new rounding method. There is a remarkable generic procedure~\cite{Rag08,Raghavendra09howto,RS09,Aus10} to devise for certain constraint satisfaction problems (CSPs) a higher-dimensional rounding method that has the best possible polynomial-time performance  unless the Unique Games Conjecture~\cite{Kho02} has a negative answer.

Despite the overall usefulness of the above ``first bit rounding'' (a.k.a. hyperplane rounding, after preprocessing), as well as its optimality in important settings such as MAXCUT~\cite{FS02,KKMO07}, combined with the fact that sharp rounding methods have been proven to exist for a large family of CSPs, there is great need to devise further rounding methods (and ways to analyse them) because there are important continuous relaxations of combinatorial optimization problems for which the aforementioned results do not shed light on the best-possible value of the associated integrality gap; understanding the Grothendieck constant is a prime example of a longstanding mystery of this nature. Of course, there is no reason to expect that such constants could be expressed in terms of elementary functions, but one might hope to find enlightening characterizations of them, better estimates, or to ``merely'' obtain a versatile rounding method that could be used heuristically  in practice. More ways to reason about the rounding problem are needed also because the aforementioned generic rounding methods for CSPs are sharp (assuming the Unique Games Conjecture) only when the integrality gap is $O(1)$ as $n\to \infty$, while there are  important questions (e.g.~Sparset Cut) where the integrality gap is unbounded as a function of $n$, so the pertinent issue becomes to devise methods that could be analysed so as to yield sharp asymptotics.

Our preprint~\cite{EN14} considers the following general rounding method that we call {\em Krivine diffusions}, in reference to the power of Krivine rounding schemes that was established in~\cite{BMMN-grothendieck,NR14}. Note, however, that while the works~\cite{BMMN-grothendieck,NR14} build on a key idea that Krivine introduced in~\cite{Kri77}, their success relies on the usefulness of degrees of freedom that Krivine expected~\cite{Kri77} {\em not to be useful}. Nonetheless, Krivine's approach is the fundamental inspiration for these ideas, thus justifying naming the method in homage to his contribution. The ensuing framework allows one to bring in tools from stochastic calculus to analyse the rounding question, and in~\cite{EN14} this is shown to relate integrality gaps to certain variational questions in partial differential equations. The present text, however, demonstrates how in the special setting of MAXCUT, an explicit sharp closed-form solution could be obtained.

\subsection{Krivine diffusions} Consider the following method to associate to $u\in \R^d$ a random sign $\sigma_u\in \{-1,1\}$. We call any such procedure a discrete-time Krivine diffusion. We will pass later to a natural continuous-time limit in order to use analytic tools to understand the behavior of the rounding procedure.

\begin{itemize}
\item{\bf Datum.} Each instance of the method is determined by an integer $T\in \N$, a function $f:\R\to \R$ and for every $t\in \{1,\ldots,T\}$ a function $F_t:\R^{t}\to \R$.
\item{\bf Randomness.} Let $\gamma_1,\ldots\gamma_T$ be independent standard Gaussian random variables in $\R^d$, i.e., their density at $(x_1,\ldots,x_d)\in \R^d$ is proportional to $\exp(-(x_1^2+\ldots+x_d^2)/2)$.
\item{\bf Evolution.} For every $u\in \R^d$ consider the random numbers $X_u(0),\ldots,X_u(T)\in \R$ that are defined by setting $X_u(0)=0$, and proceeding  recursively by setting
\begin{equation}\label{eq:def X process}
 X_u (t+1 ) \eqdef X_u (t ) + F_{t} \big(X_u (1 ), \dots,  X_u (t ) \big) f\big( \langle \gamma_t, u\rangle \big).
\end{equation}
\item{\bf Output.} For each $u\in \R^d$, if there is $t\in \{1,\ldots,T\}$ with $|X_u(t)|\ge 1$, then let $T_u$ be the first time that this occurs, i.e.,  $T_u=\min\{t\in \{1,\ldots,T\}:\  |X_u(t)|\ge 1\}$, and  output   $\sigma_u=\sign(X_u(T_u))$. Otherwise, namely if $|X_u(t)|<1$ for all  $t\in \{1,\ldots,T\}$, then let $\sigma_u$ be uniformly distributed over $\{-1,1\}$.
\end{itemize}

In other words, the datum of a Krivine diffusion consists of the recursion coefficients in~\eqref{eq:def X process}  and the number $T$ of recursion steps. With this given, one chooses $T$ independent Gaussians and evolves the random numbers $\{X_u(t)\}_{t=0}^T$ according to the update rule~\eqref{eq:def X process} in which the increment $X_u(t+1)-X_u(t)$ is proportional to a function of the random dot-product $\langle \gamma_t, u\rangle$, where the proportion factor may depend on the values $X_u(1),\ldots,X_u(t)$ that were determined thus far. The output $\sigma_u\in \{-1,1\}$ is the sign of this process at the first time that it exists the interval $(-1,1)$ if such a time exists, and otherwise it is a uniformly random element of $\{-1,1\}$. Note that the same Gaussians $\gamma_1,\ldots,\gamma_T$ are used here for every $u\in \R^d$, thus inducing an important and subtle correlation between the signs $\{\sigma_u\}_{u\in \R^d}$.

There are degrees of freedom that the above construction did not exploit. One obvious variant is to allow the recursion coefficient $f( \langle \gamma_t, u\rangle)$ in~\eqref{eq:def X process} to have a further dependence on $t$ of the form $f_t( \langle \gamma_t, u\rangle)$ for some $f_t:\R\to \R$. We dropped this flexibility because thus far we did not find any use for it.

A more substantial variant is to consider distributions over families of Krivine diffusions. This will not be needed in the present note, but it is worthwhile to mention in passing that distributions over {\em pairs of} Krivine diffusions are used in the following theorem from~\cite{EN14}.

\begin{theorem}[Krivine diffusions attain the Grothendieck constant]
For any $\eps > 0$, there exists $T\in \N$ and a distribution $\mu$ over pairs of Krivine diffusions $\{(X_u(t),Y_u(t))\}_{t=0}^T$
such that for every $m,n\in \N$, every $u_1,\ldots,u_m,v_1,\ldots,v_n\in \SS^{d-1}$ and every $m$ by $n$ matrix $\A=(a_{ij})\in \mathsf{M}_{m\times n}(\R)$,
\begin{equation}\label{eq:NR Gro}\sum_{i=1}^n\sum_{j=1}^n a_{ij}\langle u_i,v_j\rangle \le (K_G+\eps) \sum_{i=1}^n\sum_{j=1}^n a_{ij}\EE \big[\sign\big(X_u(T)\big)\sign\big(Y_u(T)\big)\big]+\eps\sum_{i=1}^n\sum_{j=1}^n |a_{ij}|,
\end{equation}
where $K_G$ is the Grothendieck constant and the expectation in~\eqref{eq:NR Gro} is with respect to both $\mu$ and the underlying randomness of the Krivine diffusions, namely over the Gaussians $\gamma_1,\ldots,\gamma_T$.
\end{theorem}

A much more well-understood rounding problem concerns the special case of~\eqref{eq:NR Gro} in which the coefficients $\{a_{ij}:\ (i,j)\in \n^2\}$ are non-positive. An especially important instance of this is the MAXCUT problem; the lack of cancellations between the summands in the setting of MAXCUT allows for the success of understanding it fully assuming the Unique Games Conjecture~\cite{GW95,KKMO07}.

Abbasi-Zadeh, Bansal, Guruganesh, Nikolov, Schwartz and Singh studied in~\cite{ABGNSS18} a special case of Krivine diffusions (which they discovered independently of~\cite{EN14}), called \emph{sticky Brownian motion} (and a ``slowed down'' version thereof). It is shown in~\cite{ABGNSS18} that the integrality gap for MAXCUT  that can be attained by sticky Brownian motion is quite close to the Goemans--Williamson approximation ratio~\cite{GW95}, relying on  analytic reasoning and computer-assisted numerical computations. One of the main open questions that were posed in~\cite{ABGNSS18}  is whether the Goemans--Williamson approximation ratio can be achieved exactly using this approach. Here we prove that this is indeed the case.

\subsection{The continuous setting} We begin by recalling the slowed-down sticky Brownian motion rounding scheme of~\cite{ABGNSS18}. Its input  is  a sequence of unit vectors $u_1,\ldots,u_n\in \RR^d$ and its goal is to produce signs $\sigma_1,\ldots,\sigma_n \in \{-1,1\}$. To do so, we will use below basic facts from stochastic calculus, all of which are standard and appear in introductory texts such as~\cite{Oks03,KS91}. Let $\{B(t)\}_{t\ge 0}$ be a Brownian motion in $\RR^d$ adapted to the filtration $\{\FF_t\}_{t\ge 0}$. Sticky Brownian motion refers to the associated random signs
\begin{equation*}\label{eq:defW}
\forall\, i\in \n,\qquad \sigma_i\eqdef \langle u_i,  B(T_i)\rangle,\qquad\mathrm{where}\qquad  T_i\eqdef \min\{t\ge 0:\ |\langle u_i,  B(t)\rangle|=1\}.
\end{equation*}
Thus, the sign $\sigma_i$ is extracted when the stochastic process $\{\langle u_i,  B(t)\rangle\}_{t\ge 0}$ first reaches the set $\{-1,1\}$; hence the terminology ''sticky.'' Next, consider the following modification of this procedure. Fix a continuous function $\varphi : [-1,1] \to [0, \infty)$ that satisfies
\begin{equation}\label{eq:phi limit assumption}
\lim_{s\to 1^-}\varphi(s)=\lim_{s\to -1^+}\varphi(s)=0\qquad\mathrm{and}\qquad \forall\, s\in (-1,1),\qquad \varphi(s)>0.
\end{equation}
Let $\{W_u^\varphi(t)\}_{(u,t)\in \R^d\times [0,\infty)}$ be a stochastic process that satisfies the stochastic differential equation
\begin{equation}\label{eq:defWS}
\forall(u,t)\in \R^d\times [0,\infty),\qquad W_u^\varphi(0) = 0,\qquad\mathrm{and}\qquad  \ud W_u^\varphi(t) = \varphi\big(W_u^\varphi(t)\big) \langle u, \ud B(t)\rangle.
\end{equation}
 The existence and uniqueness (in distribution) of $\{W_u^\varphi(t)\}_{(u,t)\in \R^d\times [0,\infty)}$ follows from the rudimentary theory of stochastic differential equations; thorough treatments of this appear in e.g.~\cite{KS91,Oks03}. In particular, $\{W_u^\varphi(t)\}_{t\ge 0}$ is a martingale (with respect to the Brownian filtration $\{\FF_t\}_{t\ge 0}$) for every $u\in \R^d$, and by the martingale convergence theorem the following limit exists almost-surely.
$$
\sigma^\varphi_u\eqdef \lim_{t\to\infty} W_u^\varphi(t).
$$
The assumptions~\eqref{eq:phi limit assumption} on $\varphi$ ensure that $\sigma^\varphi_u\in \{-1,1\}$ almost-surely. Since the speed $\varphi$ satisfies $\varphi(s) \to 0$ as $s \to \{\pm 1\}$, this process is called ``slowed-down'' sticky Brownian motion.

\begin{remark} For each fixed $u\in \R^d$ the equation~\eqref{eq:defWS}  that governs the evolution of the process $\{W_u^\varphi(t)\}_{\t\ge 0}$ is a one-dimensional stochastic differential equation. One can solve~\eqref{eq:defWS} separately for each $u\in \R^d$, but it is important to stress that the resulting rounding procedure relies on the fact that the underlying Brownian motion that is used to drive this family of  stochastic differential equations is the same for all $u\in \R^d$, thus inducing important correlations which are exploited crucially both below and in~\cite{EN14,ABGNSS18}.
\end{remark}

A mechanical approximation argument verifies that the above processes can be obtained  as continuous limits of Krivine diffusions; we will not include the straightforward details here (they are meaningful for understanding the context, and also a discretization is needed for efficient implementation of the rounding method, but such issues are not needed for the ensuing proof).

The work~\cite{ABGNSS18} makes the choice $\varphi(s) = (1-s^2)^{\alpha}$ for $\alpha>0$. Here we pinpoint the following different choice which recovers the Goemans--Williamson approximation ratio and thus answers positively the question raised in~\cite{ABGNSS18}. Define $\xi:[-1,1]\to [0,\infty)$ by setting
\begin{equation}\label{eq:def xi}
\forall\, s\in [-1,1],\qquad \xi(s) \eqdef \frac{\sqrt{2}}{\sqrt{\pi}} e^{ - \frac12 \Phi^{-1}\big(\frac{1-s}{2}\big)^2  },
\end{equation}
where $\Phi:\R\to \R$ is the standard Gaussian cumulative distribution function, i.e.,
\begin{equation}\label{eq:id}
\forall\,x\in \R,\qquad \Phi(x)\eqdef \frac{1}{\sqrt{2\pi}}\int_\infty^x e^{-\frac{s^2}{2}}\ud s,\qquad\mathrm{thus}\qquad
\Phi''( x ) = - x \Phi'(x).
\end{equation}
With the following theorem at hand, one concludes identically to~\cite{GW95} that the slowed-down sticky Brownian motion with speed $\xi$ attains the Goemans--Williamson approximation ratio  for MAXCUT.

\begin{theorem} \label{thm:GW} For $u\in \R^d$ and $t\ge 0$ write $W_u^\xi(t)=W_u(t)$ and $\sigma_u^\xi=\sigma_u\in \{-1,1\}$, where $\xi$ is given in~\eqref{eq:def xi}. Then,
\begin{equation}\label{eq:GW}
\forall\, u,v\in \mathbb{S}^{d-1},\qquad \EE \big[\sigma_u \sigma_v \big] = \frac{2}{\pi} \arcsin(\langle u,v\rangle).
\end{equation}
\end{theorem}

\section{Proof of Theorem~\ref{thm:GW}}

Let $\{B(t)\}_{t\ge 0}$ be a Brownian motion in $\RR^d$ adapted to the filtration $\{\FF_t\}_{t\ge 0}$. Consider the process
\begin{equation}\label{eq:def Z}\forall (x,t)\in \R^d\times [0,\infty),\qquad Z_x(t) \eqdef \int_0^t e^{-\frac{s}{2}} \langle x, \ud B(s) \rangle.
\end{equation}
The convergence of the integral $\int_0^\infty e^{-s} ds$ implies that  the following limit  exists almost-surely.
\begin{equation}\label{eq:def Zu}
Z_x(\infty) \eqdef \lim_{t \to \infty} Z_w(t)=\int_0^\infty e^{-\frac{s}{2}} \langle x, \ud B(s) \rangle.
\end{equation}

Fix $u,v\in \mathbb{S}^{d-1}$. Because $u$ and $v$ are unit vectors and $\int_0^\infty e^{-s}\ud s=1$, it follows from~\eqref{eq:def Zu} that both $Z_u(\infty)$ and $Z_v(\infty)$ are standard Gaussian random variables, and their   covariance equals
$$
\EE\big[Z_u(\infty)Z_v(\infty)\big] \stackrel{\eqref{eq:def Zu}}{=} \int_0^\infty e^{-s} \ud [\langle u, B(s) \rangle \langle v, B(s) \rangle ]_s = \int_0^\infty e^{-s}\langle u,v\rangle\ud s=\langle u,v\rangle.
$$
Hence, by the original computation  of Goemans and Wiliamson~\cite{GW95} (using  elementary planar geometry), we have the following equality (Grothendieck's identity~\cite{Gro53,LP68}) .
\begin{equation}\label{eq:standard GW computation}
\EE \big[ \mathrm{sign} \big (Z_u(\infty)\big) \sign\big(Z_v(\infty) \big) \big] = \frac{2}{\pi} \arcsin(\langle u,v\rangle).
\end{equation}

Due to~\eqref{eq:standard GW computation},  Theorem~\ref{thm:GW} will be proven if we show  that $(\sigma_u,\sigma_v)$ and $(\sign(Z_u(\infty)),\sign(Z_v(\infty)))$ have the same distribution.  More generally, by working  with continuous modifications  of  $\{W_x(t)\}_{(x,t)\in \mathbb{S}^{d-1}\times [0,\infty)}$ and $\{Z_x(t)\}_{(x,t)\in \mathbb{S}^{d-1}\times [0,\infty)}$ (as we may, see e.g.~\cite{Kun84}), the ensuing argument establishes that the sign-valued processes $\{\sigma_x\}_{x\in\mathbb{S}^{d-1}}$ and $\{\sign(Z_x(\infty))\}_{x\in \mathbb{S}^{d-1}}$ are equal almost-surely.

For every $u\in \R^d$, consider the martingale that is defined by
\begin{equation}\label{eq:def M}
\forall\, t\ge 0,
\qquad  M_u(t) \eqdef \EE\left[ \mathrm{sign} \big( Z_u(\infty)\big)  \middle| \FF_t \right].
\end{equation}
By the martingale convergence theorem, we almost-surely have
\begin{equation}\label{eq:martingale convergence}
\lim_{t\to\infty} M_u(t)\stackrel{\eqref{eq:def M}}{=}\EE\left[ \mathrm{sign} \big( Z_u(\infty)\big)  \middle| \FF_\infty \right]=\mathrm{sign} \big( Z_u(\infty)\big).\end{equation}
We will show below that the stochastic processes $\{M_u(t)\}_{u\in \mathbb{S}^{d-1}\times [0,\infty)}$ and $\{W_u(t)\}_{u\in \mathbb{S}^{d-1}\times [0,\infty)}$ are equal. Since, by definition, $\sigma_u=\lim_{t\to \infty}W_u(t)$ almost-surely, together with~\eqref{eq:martingale convergence} this would imply the desired equi-distribution of $(\sigma_u,\sigma_v)$ and $(\sign(Z_u(\infty)),\sign(Z_v(\infty)))$.

Note that $\int_t^\infty e^{-s/2} \langle u, \ud B(s) \rangle$ is independent of $Z_u(t)$ and has the same distribution as   $e^{-t/2}\gamma$, where $\gamma$ is a standard Gaussian random variable that is independent of $\{Z_u(t)\}_{(u,t)\in \R^d\times [0,\infty]}$. Hence,
\begin{equation}\label{eq:M identity}
M_u(t) \stackrel{\eqref{eq:def M}}{=}2 \PP\left[ Z_u(\infty)\phantom{\big(}\ge 0  \middle| \FF_t \right] - 1=2 \PP\big[Z_u(t)+e^{-\frac{t}{2}}\gamma>0 \big] - 1= 1 - 2 \Phi\big( - e^{\frac{t}{2}} Z_u(t) \big).
\end{equation}
It\^o's formula yields,
\begin{align}\label{eq:M equation}
\begin{split}
\ud M_u(t) & = 2 \Phi'\big( - e^{\frac{t}{2}} Z_u(t) \big) e^{\frac{t}{2}} \ud Z_u(t)
-\Phi''\big(- e^{\frac{t}{2}} Z_u(t)\big) \ud t + e^{\frac{t}{2}} Z_u(t) \Phi'\big(- e^{\frac{t}{2}} Z_u(t) \big) \ud t \\
& \stackrel{\ \, \,\eqref{eq:id}\wedge\eqref{eq:M identity}}{=} 2 \Phi' \left ( \Phi^{-1} \Big ( \frac{1-M_u(t)}{2} \Big ) \right ) e^{\frac{t}{2}} \ud Z_u(t)
 \stackrel{\ \,\eqref{eq:def xi}\wedge \eqref{eq:def Z}}{=} \xi \big(M_u(t)\big) \langle u, \ud B(t) \rangle.
\end{split}
\end{align}
Recalling that $\{W_u(t)\}_{u\in \mathbb{S}^{d-1}\times [0,\infty)}$  satisfies the stochastic differential equation~\eqref{eq:defWS} with $\varphi=\xi$, we see from~\eqref{eq:M equation} that the processes  $\{M_u(t)\}_{u\in \mathbb{S}^{d-1}\times [0,\infty)}$ and $\{W_u(t)\}_{u\in \mathbb{S}^{d-1}\times [0,\infty)}$ satisfy the same stochastic differential equation, so by It\^o's uniqueness theorem  $W_u(t) = M_u(t)$ for all $t \in [0,\infty)$ and all $u\in \mathbb{S}^{d-1}$. \qed

\bibliographystyle{alphaabbrvprelim}
\bibliography{bib-kri}

\newcommand{\etalchar}[1]{$^{#1}$}
\begin{thebibliography}{AZBG{\etalchar{+}}18}
\expandafter\ifx\csname urlstyle\endcsname\relax
  \providecommand{\doi}[1]{doi:\discretionary{}{}{}#1}\else
  \providecommand{\doi}{doi:\discretionary{}{}{}\begingroup
  \urlstyle{rm}\Url}\fi

\bibitem[Aus10]{Aus10}
P.~Austrin.
\newblock Towards sharp inapproximability for any 2-{CSP}.
\newblock \emph{SIAM J. Comput.}, 39(6):2430--2463, 2010.

\bibitem[AZBG{\etalchar{+}}18]{ABGNSS18}
S.~Abbasi-Zadeh, N.~Bansal, G.~Guruganesh, A.~Nikolov, R.~Schwartz, and
  M.~Singh.
\newblock Sticky brownian rounding and its applications to constraint
  satisfaction problems, 2018.
\newblock ArXiv:1812.07769.

\bibitem[BMMN13]{BMMN-grothendieck}
M.~Braverman, K.~Makarychev, Y.~Makarychev, and A.~Naor.
\newblock The {G}rothendieck constant is strictly smaller than {K}rivine's
  bound.
\newblock \emph{Forum Math. Pi}, 1:e4, 42, 2013.

\bibitem[EN14]{EN14}
R.~Eldan and A.~Naor.
\newblock Krivine diffusions, 2014.
\newblock Preprint.

\bibitem[FS02]{FS02}
U.~Feige and G.~Schechtman.
\newblock On the optimality of the random hyperplane rounding technique for
  {MAX} {CUT}.
\newblock \emph{Random Structures Algorithms}, 20(3):403--440, 2002.
\newblock Probabilistic methods in combinatorial optimization.

\bibitem[Gro53]{Gro53}
A.~Grothendieck.
\newblock R\'{e}sum\'{e} de la th\'{e}orie m\'{e}trique des produits tensoriels
  topologiques.
\newblock \emph{Bol. Soc. Mat. S\~{a}o Paulo}, 8:1--79, 1953.

\bibitem[GW95]{GW95}
M.~X. Goemans and D.~P. Williamson.
\newblock Improved approximation algorithms for maximum cut and satisfiability
  problems using semidefinite programming.
\newblock \emph{J. Assoc. Comput. Mach.}, 42(6):1115--1145, 1995.

\bibitem[Kho02]{Kho02}
S.~Khot.
\newblock On the power of unique 2-prover 1-round games.
\newblock In \emph{Proceedings of the {T}hirty-{F}ourth {A}nnual {ACM}
  {S}ymposium on {T}heory of {C}omputing}, pages 767--775. ACM, New York, 2002.
\newblock \doi{10.1145/509907.510017}.

\bibitem[KKMO07]{KKMO07}
S.~Khot, G.~Kindler, E.~Mossel, and R.~O'Donnell.
\newblock Optimal inapproximability results for {MAX}-{CUT} and other
  2-variable {CSP}s?
\newblock \emph{SIAM J. Comput.}, 37(1):319--357, 2007.

\bibitem[Kri77]{Kri77}
J.-L. Krivine.
\newblock Sur la constante de {G}rothendieck.
\newblock \emph{C. R. Acad. Sci. Paris S\'{e}r. A-B}, 284(8):A445--A446, 1977.

\bibitem[KS91]{KS91}
I.~Karatzas and S.~E. Shreve.
\newblock \emph{Brownian motion and stochastic calculus}, volume 113 of
  \emph{Graduate Texts in Mathematics}.
\newblock Springer-Verlag, New York, second edition, 1991.
\newblock ISBN 0-387-97655-8.
\newblock \doi{10.1007/978-1-4612-0949-2}.

\bibitem[Kun84]{Kun84}
H.~Kunita.
\newblock Stochastic differential equations and stochastic flows of
  diffeomorphisms.
\newblock In \emph{\'{E}cole d'\'{e}t\'{e} de probabilit\'{e}s de
  {S}aint-{F}lour, {XII}---1982}, volume 1097 of \emph{Lecture Notes in Math.},
  pages 143--303. Springer, Berlin, 1984.
\newblock \doi{10.1007/BFb0099433}.

\bibitem[LP68]{LP68}
J.~Lindenstrauss and A.~Pe{\l}czy\'{n}ski.
\newblock Absolutely summing operators in {$L_{p}$}-spaces and their
  applications.
\newblock \emph{Studia Math.}, 29:275--326, 1968.

\bibitem[NR14]{NR14}
A.~Naor and O.~Regev.
\newblock Krivine schemes are optimal.
\newblock \emph{Proc. Amer. Math. Soc.}, 142(12):4315--4320, 2014.

\bibitem[Ok03]{Oks03}
B.~\O~ksendal.
\newblock \emph{Stochastic differential equations}.
\newblock Universitext. Springer-Verlag, Berlin, sixth edition, 2003.
\newblock ISBN 3-540-04758-1.
\newblock \doi{10.1007/978-3-642-14394-6}.
\newblock An introduction with applications.

\bibitem[Pis12]{Pis12}
G.~Pisier.
\newblock Grothendieck's theorem, past and present.
\newblock \emph{Bull. Amer. Math. Soc. (N.S.)}, 49(2):237--323, 2012.

\bibitem[Rag08]{Rag08}
P.~Raghavendra.
\newblock Optimal algorithms and inapproximability results for every {CSP}?
  [extended abstract].
\newblock In \emph{S{TOC}'08}, pages 245--254. ACM, New York, 2008.
\newblock \doi{10.1145/1374376.1374414}.

\bibitem[RS09a]{Raghavendra09howto}
P.~Raghavendra and D.~Steurer.
\newblock How to round any csp.
\newblock In \emph{In Proc. 50th IEEE Symp. on Foundations of Comp. Sci}. 2009.

\bibitem[RS09b]{RS09}
P.~Raghavendra and D.~Steurer.
\newblock Towards computing the {G}rothendieck constant.
\newblock In \emph{Proceedings of the {T}wentieth {A}nnual {ACM}-{SIAM}
  {S}ymposium on {D}iscrete {A}lgorithms}, pages 525--534. SIAM, Philadelphia,
  PA, 2009.

\bibitem[Tsi85]{Tsi85}
B.~S. Tsirelson.
\newblock Quantum analogues of {B}ell's inequalities. {T}he case of two
  spatially divided domains.
\newblock \emph{Zap. Nauchn. Sem. Leningrad. Otdel. Mat. Inst. Steklov.
  (LOMI)}, 142:174--194, 200, 1985.
\newblock Problems of the theory of probability distributions, IX.

\end{thebibliography}

\end{document}